\documentclass[aps,nofootinbib,twocolumn,prl,showpacs,class-pre]{revtex4}
\usepackage{setspace} \usepackage{amsmath,amssymb,amsfonts,amsthm}
\usepackage{graphicx}

\newcommand{\be}{\begin{equation}}
\newcommand{\ee}{\end{equation}}
\newcommand{\beq}{\begin{eqnarray}}
\newcommand{\eeq}{\end{eqnarray}}

\begin{document}
\title{Noncommutative Geometry Spectral Action as a framework for
  unification:\\ Introduction and phenomenological/cosmological
  consequences} \author{Mairi
  Sakellariadou~\footnote{mairi.sakellariadou@kcl.ac.uk}}
\affiliation{Department of Physics, King's College, University of
  London, Strand WC2R 2LS, London, U.K.}

\begin{abstract}
I will summarize Noncommutative Geometry Spectral Action, an elegant
geometrical model valid at unification scale, which offers a purely
gravitational explanation of the Standard Model, the most successful
phenomenological model of particle physics. Noncommutative geometry
states that close to the Planck energy scale, space-time has a fine
structure and proposes that it is given as the product of a
four-dimensional continuum compact Riemaniann manifold by a tiny
discrete finite noncommutative space.  The spectral action principle,
a universal action functional on spectral triples which depends only
on the spectrum of the Dirac operator, applied to this almost
commutative product geometry, leads to the full Standard Model,
including neutrino mixing which has Majorana mass terms and a see-saw
mechanism, minimally coupled to gravity. It also makes various
predictions at unification scale.  I will review some of the
phenomenological and cosmological consequences of this beautiful and
purely geometrical approach to unification.

\end{abstract}


\maketitle

\section{Preface}
It is with great pleasure that I am writing this contribution for {\sl
  Mario Castagnino's Festschrift}. I have met Mario at the beginning
of my career, almost twenty years ago, and I have decided to write
here on a subject which captured my interest only very recently. In
what follows, I will summarize NonCommutative Geometry Spectral
Action, and then discuss shortly some of its phenomenological and
cosmological consequences. Since Mario has worked in various
mathematical as well as physical problems, I hope that my choice for
this contribution is the right one.

\section{Introduction}
One of the fundamental questions in theoretical physics is the origin
of space-time, an issue which is closely related to unifying all
fundamental interactions including gravity, and as such is not
independent of solving the issue of Quantum Gravity.  The motivation
of studying NonCommutative Geometry (NCG) is at least
two-fold. Namely, besides the mathematical beauty of the NCG theory,
this approach leads to a variety of cosmological and high energy
physics consequences. Let me be a bit more precise.  As far as
Cosmology is concerned, early universe cosmological models can be
tested with very accurate astrophysical data, while high energy
experiments, and in particular the Large Hadron Collider (LHC), will
test some of the theoretical pillars of these models. However, despite
the golden era of cosmology, issues like the origin of dark energy,
the hunting for the successful dark matter candidate, or the search
for a natural and well-motivated inflationary model, are still
awaiting for a definite answer. The main approaches upon which
cosmological models have been built is either String Theory or Loop
Quantum Gravity~\footnote{Cosmological models have been also, to a
  less extent, built upon other variants of Quantum Gravity, like Spin
  Foams, Wheeler-De Witt equation, Causal Dynamical Triangulations, or
  Causal Sets.}. Here, we will use NCG Spectral Action.  As High
Energy Physics is concerned, one may argue that the distinct feature
between the diffeomorphism invariance (outer automorphism) which
governs General Relativity, and the local gauge invariance (inner
automorphism) which governs gauge symmetries, may be at the origin of
the unsuccessful search for a unified theory of all interactions
including gravity. Moreover, there is a whole list of questions within
the realm of particle physics, awaiting for a definite answer from the
successful unified theory of all interactions. In particular, there is
no conceptual justification for the choice of the gauge group G$_{\rm
  SM}$=SU(3)$\times$SU(2)$\times$U(1) of the Standard Model (SM) of
electroweak and strong interactions, nor is any reason for the various
representations for the fermions and bosons in the construction of the
SM.

Following the spirit of NCG~\cite{ncg-book1, ncg-book2}, one may ague
that much below Planck scale, gravity can be safely considered as a
classical theory, while as energies approach Planck scale, the quantum
nature of space-time reveals itself, and the naive approach, valid
however at low energy scales, that physics can be described by the sum
of the Einstein-Hilbert and the SM action becomes just an invalid
approximation. To guess the appropriate structure of space-time at
Planckian energy scales is a rather dangerous issue, and one is often
making an extrapolation by (too) many orders of magnitude. In this
sense, NCG Spectral Action follows a rather safer approach, namely it
looks for a hidden structure in the functional of gravity coupled to
the SM at today's (low) energy scales.

In what follows, we will adopt an effective theory approach and
consider the simplest case beyond commutative spaces, namely we will
suppose that at energies below but close to the Planck scale,
space-time is almost commutative. At higher energy scales, space-time
should become noncommutative in a nontrivial way, while at energies
above the Planck scale the whole concept of geometry may altogether
become meaningless.

%%%%%%%%%%%%%%%%%%%%%%%%%%%%%%%%%%%%%%%%%%%%%%%%%%%%%%%%%%%%%%%%%%%%%
\section {NCG Spectral Action}
%%%%
The basic idea of the NCG Spectral Action approach, based upon three
{\sl ansatz}, is that the Standard Model of electroweak and strong
interactions should be seen as a phenomenological model, which however
dictates the appropriate geometry of space-time, so that the
Maxwell-Dirac action functional leads to the SM action.  This proposal
then implies that the geometrical space should be given by the tensor
product ${\cal M}\times{\cal F}$ of a continuum compact~\footnote{The
  Euclidean space-time manifold is taken to be compact for
  simplicity.} Riemannian manifold ${\cal M}$ (geometry for
space-time) and a tiny discrete finite noncommutative space ${\cal F}$
(internal geometry for the SM) composed of just two points. The finite
geometry ${\cal F}$ will be chosen so that it is one of the simplest
and most natural finite noncommutative geometries of the right
dimension to solve the fermion doubling problem.  The choice of ${\cal
  M}\times{\cal F}$ as the appropriate geometrical space consists the
first {\sl ansatz}.

The noncommutative nature of the discrete space ${\cal F}$ is given by
a spectral triple~\footnote{Applying the definition of spectral type
  in the context of noncommutative geometry, it must be seen as
  setting a unitary Hilbert space representation of a {\sl setup} that
  allows to manipulate algebraically coordinates and measure
  distances.}  $({\cal A, H, D})$, where ${\cal A}$ is an involution
of operators on the finite-dimensional Hilbert space ${\cal H}$ of
Euclidean fermions, and ${\cal D}$ is a self-adjoint unbounded
operator in ${\cal H}$, such that $J{\cal D} = \epsilon'{\cal D}J$,
where $J$ is an antilinear isometry of the finite dimensional Hilbert
space ${\cal H}$, $J:{\cal H}\rightarrow{\cal H}$, with the properties
\be
J^2=\epsilon~~,~~J\gamma=\epsilon''\gamma J~,
\ee
where $\gamma$ is the chirality operator and
$\epsilon,\epsilon',\epsilon''\in \{\pm 1\}^3$.  Following the spirit
of NCG spectral action, the SM minimally coupled with Einstein gravity
appears~\cite{ccm} naturally as pure gravity on the ${\cal
  M}\times{\cal F}$ space.

Let me first discuss the distinction between the metric (or spectral)
dimension, given by the behavior of the eigenvalues of the Dirac
operator, and the K-theoretic dimension, an algebraic dimension based
on K-theory. I will start with the metric dimension: Since the
relevant Dirac operator for space-time is the ordinary Dirac operator
on a curved space-time, the metric dimension is equal to four. The
internal Dirac operator consists of the fermionic mass matrix, which
has a finite number of eigenvalues, and therefore the internal metric
dimension is equal to zero.  As a result, the metric dimension of the
${\cal M}\times{\cal F}$ geometry is just four, the same as that of
the ordinary space-time manifold. I will now discuss the K-theoretic
dimension: There are 8 possible combinations for
$\epsilon,\epsilon',\epsilon''$ and this defines a K-theoretic
dimension of noncommutative space modulo 8. To resolve the fermion
doubling problem, by projecting out the unphysical degrees of freedom
resting in the internal space, the real structure of the finite
geometry ${\cal F}$ turns out to be such that its K-theoretic
dimension is equal to six~\cite{ac2006,fdp}, and in this case
$(\epsilon, \epsilon', \epsilon'')=(1,1,-1)$. Thus, the K-theoretic
dimension of the product space $\mathcal{M}\times {\cal F}$ is equal
to $10\sim 2\ {\rm modulo}\ 8$, allowing one to impose simultaneously
the reality and Weyl conditions in the Minkowskian continued forms.

It becomes then clear that the reason for introducing ${\cal F}$ is to
correct the K-theoretic dimension from four to ten (modulo 8). In
other words, the fermion doubling problem requires~\cite{ac2006,fdp}
crossing the ordinary four-dimensional continuum by a space of
K-theoretic dimension 6. Classifying all irreducible finite
noncommutative geometries of K-theoretic dimension six, it was
shown~\cite{cc0706} that the dimension (per generation) is a square of
an integer $k$.

Let us now go back to the discussion about the spectral triple $({\cal
  A, H, D})$. The algebra ${\cal A}$, related to the gauge group of
local gauge transformations, is the algebra of coordinates. Within
NCG, all information about a space is encoded in the algebra of
coordinates ${\cal A}$.  By assuming that the algebra constructed in
the product geometry ${\cal M}\times {\cal F}$ is symplectic-unitary,
${\cal A}$ must be of the form~\cite{Chamseddine:2007ia}
\begin{equation}
\mathcal{A}=M_{a}(\mathbb{H})\oplus M_{k}(\mathbb{C})~,
\end{equation}
with $k=2a$ and $\mathbb{H}$ being the algebra of quaternions.  The
field of quaternions $\mathbb{H}$ plays an important r\^ole in this
construction and its choice remains to be explained. To obtain the SM
we will assume quaternion linearity.

The first possible value for the even number $k$ is 2, corresponding
to a Hilbert space of four fermions; it is ruled out from the
existence of quarks. The second one, $k=4$, leads to the correct
number of $k^2=16$ fermions in each of the three generations.
Certainly if at LHC new particles are discovered, one may be able to
accommodate them by including a higher value for the even number $k$.
Notice that considering three generations ($N=3$) is a physical input
in NCG.  Certainly, as far as physics is concerned, violation of CP is
a reason for setting $N\geq 3$, but one would like to come up with
also a mathematical justification.  The choice of ${\cal A}$ is the
main input in the entire approach and its choice consists the second
{\sl ansatz}.

The choice of Hilbert space has no importance, since all separable
infinite-dimensional Hilbert spaces are isomorphic. 

The operator $D$ corresponds to the inverse of the Euclidean
propagator of fermions, and is given by the Yukawa coupling matrix
which encodes the masses of the elementary fermions and the
Kobayashi--Maskawa mixing parameters. The commutator $[D,a]$, with
$a\in {\cal A}$, plays the r\^ole of the differential quotient
$da/ds$, with $ds$ the unit of length.  The familiar geodesic formula
\be
d(x,y)={\rm inf}\int_\gamma ds~,
\ee
where the infimum is taken over all possible paths connecting $x$ to
$y$, which is used to determine the distance $d(x,y)$ between two
points $x$ and $y$ within Riemannian geometry, is replaced by
\be
d(x,y)={\rm sup}\{|f(x)-f(y)|: f\in {\cal A}, ||[D,f]|| \leq 1\}~,
\ee
where $D$ is the inverse of the line element $ds$, within the
noncommutative spectral geometry.  

The operator $D$ is assumed~\cite{ccm} to commute with the sub-algebra
${(\lambda, \lambda, 0); \lambda\in \mathbb{C}}$. The meaning of this
condition is clear as far as physics is concerned, namely it means
that the photon is massless, however a conceptual mathematical reason
for only considering metrics satisfying this requirement is lacking.

The fermions of the SM provide the Hilbert space ${\cal H}$ of a
spectral triple for the algebra ${\cal A}$, while the bosons of the
SM, including the Higgs boson, are obtained through inner fluctuations
of the Dirac operator of the product ${\cal M}\times {\cal F}$
geometry. Hence, the Higgs boson, which generates the masses of
elementary particles through Spontaneous Symmetry Breaking, becomes
just a gauge field corresponding to a finite difference.  Note that
the corresponding mass scale specifies the inverse size of the
discrete geometry ${\cal F}$.

To obtain the full Lagrangian of the SM, minimally coupled to gravity,
we will apply the Spectral Action principle, which consists the third
{\sl ansatz}. It states that the Dirac operator connects the two
pieces of the product ${\cal M}\times{\cal F}$ geometry nontrivially.
The spectral action functional~\footnote{Similar to Fourier Transform
  in commutative geometry.} $S$ depends only of the spectrum of the
Dirac operator and is of the form ${\rm Tr}(f(D/\Lambda))$, with
$\Lambda$ giving the energy scale~\footnote{The arbitrary mass scale
  in the spectral action for the Dirac operator can be made dynamical
  by introducing a dilaton field, which guarantees the scale
  invariance of the SM interactions, and provides a mechanism to
  generate mass hierarchies~\cite{cc2005-dilaton}.} and $f$ being a
cut-off function, whose choice plays only a small r\^ole. Note that
both $D$ and $\Lambda$ have physical dimensions of a mass and there is
no absolute scale on which they can be measured.

Using heat kernel methods the trace ${\rm Tr}(f(D/\Lambda))$ can be
written in terms of the geometrical Seeley-deWitt coefficients $a_n$,
as~\cite{sdw-coeff}
\be
\sum_{n=0}^\infty F_{4-n}\Lambda^{4-n}a_n~,
\ee
where the function $F$ is defined such that $F(D^2)=f(D)$.
Defining the moments
\be\label{eq:moments0}
f_k\equiv\int_0^\infty f(u) u^{k-1}{\rm d}u\ \ ,\ \ \mbox{for}\ \ k>0 ~,
\ee
and $f_0\equiv f(0)$, one finds
\beq \label{eq:moments}
 F_4&=&2f_4~,\nonumber\\ F_2&=&2f_2~,\nonumber\\ F_0&=&f_0~,\nonumber\\
F_{-2n}&=&\Big[(-1)^n\Big(\frac {\rm d}{2u{\rm d}u}\Big)^n
  f\Big](0)\ \ \mbox{for}\ \ n\geq 1~,
\eeq
while the Seeley-deWitt coefficients $a_n$ are known for any second
order elliptic differential operator. 

The spectral action can be expanded in powers of the scale $\Lambda$
in the form~\cite{ccm,ac1996,ac1997}
\begin{equation}
\label{eq:sp-act}
{\rm Tr}\left(f\left(\frac{D}{\Lambda}\right)\right)\sim 
\sum_{k\in {\rm DimSp}} f_{k} 
\Lambda^k{\int\!\!\!\!\!\!-} |D|^{-k} + f(0) \zeta_D(0)+ {\cal O}(1)~,
\end{equation}
where $f_k$ are the momenta of the function $f$ given in
Eq.~(\ref{eq:moments0}), the noncommutative integration is defined in
terms of residues of zeta functions, $\zeta_D (s) = {\rm
  Tr}(|D|^{-s})$ at poles of the zeta function, and the sum is
over points in the {\sl dimension spectrum} of the spectral triple.

Considering the Riemannian geometry to be four-dimensional,
the asymptotic expansion of the trace reads~\cite{nonpert}
\beq\label{asymp-exp}
{\rm Tr}\left(f\left(\frac{D}{\Lambda}\right)\right)\sim
2\Lambda^4f_4a_0+2\Lambda^2f_2a_2+f_0a_4+\cdots\nonumber\\
+\Lambda^{-2k}f_{-2k}a_{4+2k}+\cdots~.
\eeq
The smooth even function $f$, which decays fast at infinity, only
enters in the multiplicative factors:
\beq
f_4&=&\int_0^\infty f(u)u^3 du~,\nonumber\\
f_2&=&\int_0^\infty f(u)u du~,\nonumber\\
f_0&=&f(0)~,\nonumber\\
f_{-2k}&=&(-1)^k\frac{k!}{(2k)!} f^{(2k)}(0)~.
\eeq
%$
Since $f$ is taken as a cut-off function, its Taylor expansion at zero
vanishes, thus its asymptotic expansion, Eq.~(\ref{asymp-exp}),
reduces to just
\be
{\rm Tr}\left(f\left(\frac{D}{\Lambda}\right)\right)\sim
2\Lambda^4f_4a_0+2\Lambda^2f_2a_2+f_0a_4~.
\ee
Hence, $f$ plays a r\^ole through only its momenta $f_0, f_2, f_4$,
which are three additional real parameters in the model, physically
related to the coupling constants at unification, the gravitational
constant, and the cosmological constant.  In this four-dimensional
case, the term in $\Lambda^4$ gives a cosmological term, the term in
$\Lambda^2$ gives the Einstein-Hilbert action functional with the
physical sign for the Euclidean functional integral (provided
$f_2>0$), and the $\Lambda$-independent term yields the Yang-Mills
action for the gauge fields corresponding to the internal degrees of
freedom of the metric. The scale-independent terms in the spectral
action have conformal invariance.

The spectral action functional ${\rm Tr}(f(D/\Lambda))$,
Eq.~(\ref{eq:sp-act}), accounts only for the bosonic term; the
fermionic term can be included by adding $(1/2)\langle
J\psi,D\psi\rangle$, where $J$ is the real structure on the spectral
triple and $\psi$ is a spinor in the Hilbert space ${\cal H}$ of the
quarks and leptons. The bosonic term is sufficient when cosmology is
discussed; the fermionic part becomes essential when dealing with high
energy consequences of the NCG spectral action.

It is important to emphasize that the NCG procedure outlined above is
entirely classical, though it can {\sl a priori} be quantized. The
noncommutative spectral geometry simply provides an elegant
way in which the SM of particle physics can be produced from purely
(noncommutative) geometric information.

Another issue needs some attention. To use the formalism of spectral
triples in NCG, it is convenient to work with Euclidean (i.e the
signature is $(+,+,+,+)$) rather than Lorentzian signature.  The
discussion of phenomenological aspects of the theory relies on a Wick
rotation to imaginary time~\footnote{In the Euclidean action
  functional for gravity, the kinetic terms must have the correct sign
  so that the functional is bounded below. Since such positivity is
  spoiled by the scalar Weyl mode, one must show that all other terms
  get a positive sign~\cite{ncg-book2}.}, into the standard
(Lorentzian) signature. While sensible from the phenomenological point
of view, there exists as yet no justification on the level of the
underlying theory.

Applying the asymptotic expansion of Eq.~(\ref{eq:sp-act}) to the
spectral action of the product geometry ${\cal M}\times{\cal F}$ gives
a bosonic functional $S$ which includes cosmological terms, Riemannian
curvature terms, Higgs minimal coupling, Higgs mass terms, Higgs
quartic potential and Yang-Mills terms.  The fact that the
gravitational term includes, apart the Einstein-Hilbert action with a
cosmological term and a topological term, also a conformal gravity
term with the Weyl curvature tensor and a conformal coupling of the
Higgs field to gravity, essentially makes this model different from
the usual minimal coupling of the SM to gravity.  Another difference
is that here the coefficients of the gravitational terms depend upon the
Yukawa parameters of the particle physics content of the model.

In what follows, I will write explicitly the bosonic functional in
Euclidean signature.  One has to first perform a rescaling of the
Higgs field $\varphi$ so that the kinetic terms are normalized. To
normalize the Higgs field kinetic energy we rescale $\varphi$ to:
\be
{\bf H}=\frac{\sqrt{\mathfrak{a}f_0}}{\pi}\varphi~,
\ee
where the momentum $f_0$ is physically related to the coupling
constants at unification and the coefficient $\mathfrak{a}$, related
to the fermion and lepton masses and lepton mixing, is given by
\beq\label{eq:Ys}
 \mathfrak{a}&=&{\rm Tr} \left( Y^\star_{\left(\uparrow 1\right)}
Y_{\left(\uparrow 1\right)} +  Y^\star_{\left(\downarrow 1\right)}
Y_{\left(\downarrow 1\right)}\right.
\nonumber\\ 
&&+ \left. 3\left( 
 Y^\star_{\left(\uparrow 3\right)} Y_{\left(\uparrow 3\right)}
+  Y^\star_{\left(\downarrow 3\right)} Y_{\left(\downarrow 3\right)}
\right)\right)~,
\eeq
where the $(3\times 3)$ $Y$ matrices are used to classify the action
of the Dirac operator and give the fermion and lepton masses, as well
as lepton mixing, in the asymptotic version of the spectral
action. The $Y$ matrices are only relevant for the coupling of the
Higgs field with fermions through the dimensionless matrices
$\pi/\sqrt {\mathfrak{a}f_0} Y_x$ with $x\in \{(\uparrow\downarrow
,j)\}$. Thus, $\mathfrak{a}$ has the physical dimension of a
(mass)$^2$.

With this redefinition, the kinetic term reads
\be
\int \frac{1}{2} |D_\mu{\bf H}|^2 \sqrt{g} \ d^4x~.
\ee
The normalization of the kinetic terms imposes a relation between the
coupling constants $g_1, g_2, g_3$ at unification and the coefficient
$f_0$, namely
\be\label{relation_g-f}
{g_3^2f_0\over 2\pi^2}={1\over 4} ~~\mbox{and}~~ g_3^2=g_2^2={5\over
  3}g_1^2~. \ee
Performing an asymptotic expansion of the spectral action for the
product geometry ${\cal M}\times {\cal F}$, the bosonic
action in Euclidean signature reads~\cite{ccm}
\beq\label{eq:action1} {\cal S}^{\rm E} = \int \left(
\frac{1}{2\kappa_0^2} R + \alpha_0
C_{\mu\nu\rho\sigma}C^{\mu\nu\rho\sigma} + \gamma_0 +\tau_0 R^\star
R^\star\right.  \nonumber\\
+ \frac{1}{4}G^i_{\mu\nu}G^{\mu\nu
  i}+\frac{1}{4}F^\alpha_{\mu\nu}F^{\mu\nu\alpha}
+\frac{1}{4}B^{\mu\nu}B_{\mu\nu}\nonumber\\ 
+\frac{1}{2}|D_\mu{\bf H}|^2-\mu_0^2|{\bf H}|^2\nonumber\\ \left.
- \xi_0 R|{\bf H}|^2 +\lambda_0|{\bf H}|^4
\right) \sqrt{g} \ d^4 x~, \eeq
where 
\beq\label{bc} 
\kappa_0^2&=&\frac{12\pi^2}{96f_2\Lambda^2-f_0\mathfrak{c}}
~,\nonumber\\
\alpha_0&=&-\frac{3f_0}{10\pi^2}~,\nonumber\\ 
\gamma_0&=&\frac{1}{\pi^2}\left(48f_4\Lambda^4-f_2\Lambda^2\mathfrak{c}
+\frac{f_0}{4}\mathfrak{d}\right)~,\nonumber\\ 
\tau_0&=&\frac{11f_0}{60\pi^2}~,\nonumber\\
\mu_0^2&=&2\Lambda^2\frac{f_2}{f_0}-{\frac{\mathfrak{e}}{\mathfrak{a}}}~,
\nonumber\\
\xi_0&=&\frac{1}{12}~,\nonumber\\
\lambda_0&=&\frac{\pi^2\mathfrak{b}}{2f_0\mathfrak{a}^2}~,
\eeq
with $\mathfrak{a}$ given by Eq.~(\ref{eq:Ys}) and $\mathfrak{b},
\mathfrak{c}, \mathfrak{d}, \mathfrak{e}$ given by~\cite{ccm}
\beq\label{eq:Ys-oth}
 \mathfrak{b}&=&{\rm Tr}\left(\left( Y^\star_{\left(\uparrow 1\right)}
Y_{\left(\uparrow 1\right)}\right)^2 +  \left(Y^\star_{\left(\downarrow 1\right)}
Y_{\left(\downarrow 1\right)}\right)^2\right.
\nonumber\\ 
&&+ \left. 3\left( 
 Y^\star_{\left(\uparrow 3\right)} Y_{\left(\uparrow 3\right)}\right)^2
+  3 \left(Y^\star_{\left(\downarrow 3\right)} Y_{\left(\downarrow 3\right)}
\right)^2\right)~,\nonumber\\
 \mathfrak{c}&=&{\rm Tr}\left(Y^\star_R Y_R\right)~,\nonumber\\
 \mathfrak{d}&=&{\rm Tr}\left(\left(Y^\star_R Y_R\right)^2\right),\nonumber\\
 \mathfrak{e}&=&{\rm Tr}\left(Y^\star_R Y_RY^\star_{\left(\uparrow 1\right)}
Y_{\left(\uparrow 1\right)}\right)~,
\eeq
with $Y_{\left(\downarrow 1\right)}, Y_{\left(\uparrow 1\right)},
Y_{\left(\downarrow 3\right)}, Y_{\left(\uparrow 3\right)}$ and $Y_R$
being $(3\times 3)$ matrices, with $Y_R$ symmetric.

Using Eq.~(\ref{relation_g-f}) to replace $\sqrt{\mathfrak{a}f_0}/\pi$
by $\sqrt{\mathfrak{a}}/(g\sqrt{2})$, the notations of the Higgs fields
change to~\cite{ccm} 
\beq
{\bf H}&=&\frac{\sqrt{\mathfrak{a}}}{g\sqrt{2}}(1+\psi)~,\nonumber\\
&=&({2M\over g}+H-i\phi^0-i\sqrt{2}\phi^+)~,
\eeq
where $H, \phi^0, \phi^+$ are Higgs fields and $M$ stands for the mass
of the $W$ gauge boson.

The relations, Eq.~(\ref{bc}), above rely on the validity of the
asymptotic expansion at $\Lambda$, and are therefore tied intimately
to the scale at which the expansion is performed.  There is {\sl a
  priori} no reason for the constraints to hold at scales below the
unification scale $\Lambda$, since they represent mere boundary
conditions.

The factor $f_0$ is fixed by the canonical normalization of the
Yang-Mills terms (not included here) in terms of the common value of
the gauge coupling constants $g$ at unification,
$f_0=\pi^2/(2g^2)$. The value of $g$ at unification scale is
determined by standard renormalization group flow, i.e. it is given a
value which reproduces the correct observed coupling at low energies,
which is not unique since the gauge couplings fail to meet exactly in
the nonsupersymmetric Standard Model (or its extension by right-handed
neutrinos). The coefficients $\mathfrak{a}, \mathfrak{b},
\mathfrak{c}, \mathfrak{d}$ and $\mathfrak{e}$ are the Yukawa and
Majorana parameters subject to renormalization group flow. Finally,
the parameter $f_2$ is {\sl a priori} unconstrained in
$\mathbb{R}^*_+$.

Let me discuss the terms appearing in Eq.~(\ref{eq:action1}). The first
two terms only depend upon the Riemann curvature tensor; the first is
the Einstein-Hilbert term with the second one being the Weyl curvature
term. Thus, the first two terms are the Riemannian curvature terms;
the third one is the cosmological term.  The fourth term
\be R^\star
R^\star=\frac{1}{4}\epsilon^{\mu\nu\rho\sigma}\epsilon_{\alpha\beta\gamma\delta}
R^{\alpha\beta}_{\mu\nu}R^{\gamma\delta}_{\rho\sigma}~,\nonumber\ee
is the topological term that integrates to the Euler characteristic,
hence is nondynamical.  The three next terms are the Yang-Mills terms.
The eighth term is the scalar minimal coupling term, the next one is
the scalar mass term, and the last one is the scalar quartic potential
term. There is one more term, the $- \xi_0 R|{\bf H}|^2$, that couples
gravity with the SM. For $\xi_0=1/12$, this term encodes the
conformal coupling between the Higgs field and the Ricci
curvature. Let me sketch how this conformal coupling arises: The
coupling term between the Higgs field and the Ricci curvature,
appearing in the spectral action functional, is
$-f_0/(12\pi^2)\mathfrak{a}R|\phi|^2$, which after rescaling ${\bf
  H}=(\sqrt{\mathfrak{a}f_0}/\pi)\phi$, leads to the term $-R|{\bf
  H}|^2/12$. This shows the conformal coupling~\footnote{Conformal
  invariance is considered here solely in the matter sector; the
  Einstein-Hilbert term is not conformally invariant.} between the
background geometry and the Higgs field.  Note that the coupling term
between the Higgs field and the Ricci curvature should always be
present when one considers gravity coupled to scalar fields. The
nonminimal coupling between the Ricci curvature and the Higgs field,
which appears naturally in noncommutative spectral geometry, can have
significant consequences at high energies, such as in the early
universe~\cite{Nelson:2008uy,Nelson:2009wr,Marcolli:2009in,mmm,wjm1,wjm2}.

At this point, let me make a comment on the {\sl running} of $\xi_0$,
which will be useful later on, when I will discuss cosmological
consequences of the NCG spectral action proposal. The constraint
$\xi_0=1/12$ does not require by itself the coupling to remain
conformal~\footnote{This point is of a particular importance when one
  examines whether inflation through the Higgs field is a viable
  mechanism~\cite{Nelson:2009wr,mmm}.}, since it may run with the
energy scale.  Performing renormalization group analysis of the
nonminimally coupled SM, it was
argued~\cite{Buchbinder:1989bt,Yoon:1996yr} that there are no quantum
corrections to $\xi_0$, if it is exactly conformal at some energy
scale.  This claim was based on the observation that there are no
nonconformal values for the coupling $\xi_0$ for which there is a
renormalization group flow towards the conformal value as one runs the
SM parameters up in the energy scale. Thus, if there is an exactly
conformal coupling for the Higgs field at some specific scale, it will
be exactly conformal at all scales.

In conclusion, the spectral action, Eq.~(\ref{eq:action1}), has to be
considered as the {\sl bare action} at unification scale $\Lambda$,
where one supposes the merging of the coupling constants to take
place.  Since the NCG spectral action model lives naturally at
unification scale, it provides a suitable {\sl setup} to investigate
early universe cosmological models.  To make extrapolations to lower
energy scales one has to use Renormalization Group Equations
(RGE)~\footnote{The renormalized action will have the same form but
  with the bare quantities $\kappa_0, \alpha_0, \gamma_0, \tau_0,
  \mu_0, \xi_0, \lambda_0$, and the three gauge couplings
  $g_1,g_2,g_3$ replaced with physical quantities.}, and consider
nonperturbative effects in the spectral action.

Assuming the {\sl big desert} hypothesis, we can connect the physics
at low energies with those at $E=\Lambda$ through the standard
renormalization procedure.  This was carried out at one-loop in
Ref.~\cite{ccm}, and more recently in Ref.~\cite{Marcolli:2009in}
where Majorana mass terms for right-handed neutrinos were included and
the see-saw mechanism was taken into account.  Let me repeat that one
has to be particularly careful, since the relations given in
Eq.~(\ref{bc}) {\bf cannot} be taken as  functions of the energy scale,
this is simply incorrect; the relations in Eq.~(\ref{bc}) hold {\bf
  only} at unification scale $\Lambda$.  This leads to a major
difficulty when one would like to study lower energy astrophysical
consequences of the noncommutative spectral geometry.

Furthermore, there is another complexity. Namely, it is very difficult
to compute exactly the spectral action in its nonperturbative form,
even though recently some progress has been
achieved~\cite{nonpert}. Since the action functional ${\rm Tr}(f
(D/\Lambda))$ is not local, but its locality is only achieved when it
is replaced by the asymptotic expansion, Eq.~(\ref{asymp-exp}), one
should at least compute the next term in the asymptotic expansion, in
order to check its validity.  It has been recently
shown~\cite{nonpert} that for a space-time whose spatial sections are
3-spheres $S^3$, Wick rotated and compactified to a Euclidean model
$S^3 \times S^1$ , the spectral action can be computed explicitly in a
nonperturbative form, through the Poisson summation formula. For the
$S^3 \times S^1$, the authors of Ref.~\cite{nonpert} have demonstrated
that the spectral action is given, for any test function, by the sum
of two terms up to a remarkably tiny correction; all higher order
terms $a_{2n}$ vanish.  The authors have confirmed~\cite{nonpert}
their result by evaluating the spectral action using the heat kernel
expansion and explicitly shown that both the higher order terms $a_4$
and $a_6$ vanish.  Computations of the spectral action on other
3-manifolds, which however remain far of any realistic physical space,
has been lately presented in Ref.~\cite{mpt}.

\section{Phenomenological consequences of the NCG spectral action}

Let us assume that the function $f$ is well approximated by the
cut-off function, which then allows us to ignore higher order terms.
The NCG spectral action approach then leads to the following
phenomenological consequences~\cite{ccm}:
\\
\\
$\bullet$ Normalization of the kinetic terms dictates a relation
between the coupling constants $g_1, g_2, g_3$ and the coefficient
$f_0$, namely
\be
{g_3^2f_0\over 2\pi^2}={1\over 4} ~~\mbox{and}~~ g_3^2=g_2^2={5\over
  3}g_1^2\nonumber \ee
(i.e., Eq.~(\ref{relation_g-f})).
\\
\\
$\bullet$ It consequently implies the relation 
\be \sin^2\theta_{\rm W}=\frac{3}{8}~, \ee
which is also obtained in the context of the Grand Unified
Theories (GUTs) SU(5) and SO(10).
\\
\\
$\bullet$ The three momenta $f_0, f_2, f_4$ can be used to specify the
initial conditions on the gauge couplings, the Newton constant and the
cosmological constant, as already discussed earlier.
\\
\\
$\bullet$ Assuming the {\sl big desert} hypothesis, one can find the
running of the three couplings $\alpha_i=g_i^2/(4\pi)$.  One-loop RGE
for the running of the gauge couplings and the Newton constant, shows
that they do not meet exactly at one point, the error is though within
just few percent.  Therefore, the model does not specify a unique
unification energy. This {\sl negative} result provides useful
information about the nature of the function $f$ used in the spectral
action.  In the approach we have followed here, $f$ has been
approximated by a cut-off function for which all coefficients, given
by derivatives of $f$ at zero, of the higher order terms in the
asymptotic expansion vanish. One can therefore conclude that the
function $f$ can be safely approximated by the cut-off function,
nevertheless there exist small deviations.
\\
\\
$\bullet$ There are 16 fundamental fermions.  
\\
\\
$\bullet$ The correct representations of the fermions with respect to
the gauge group, G$_{\rm SM}$, of the SM are obtained.
\\
\\
$\bullet$ The Higgs doublet appears as part of the inner fluctuations
of the metric, and Spontaneous Symmetry Breaking mechanism arises
naturally with the negative mass term without any tuning. 
\\
\\
$\bullet$ The see-saw mechanism to give very light left-handed neutrinos
is obtained.

Moreover, the model predicts~\cite{ccm}:
\\
\\
$\bullet$ A top quark mass of $M_{\rm top}\sim 179 ~{\rm Gev}$.  
\\
\\
$\bullet$ The mass of Higgs in zeroth order approximation of the
spectral action is
\be
m_H=\sqrt{2\lambda}\frac{2M}{g}\sim 170 {\rm GeV},
\ee
with $\lambda$ the quartic Higgs coupling. This value is however ruled
out by current experimental data. Nevertheless, one should keep in
mind that the result depends on the value of gauge couplings at
unification scale, which is certainly uncertain. In addition, note
that this result was found neglecting the nonminimal coupling between
the Higgs field and the Ricci curvature.

Regarding the gravitational terms, neglecting (which is  in principle
incorrect) the nonminimal coupling between the Higgs field and the
Ricci curvature, the noncommutative spectral geometry:
\\
\\
$\bullet$ Agrees with the very weak values of the coefficients of the
quadratic curvature terms $R^{\mu\nu} R_{\mu\nu}$ and $R^2$ at low
energies, found experimentally.
\\
\\
$\bullet$ For $\Lambda\sim1.1\times 10^{17}\ {\rm GeV}$, from the
standard form of the gravitational action, $S(g)=1/(16\pi G)\int_{\rm
  M} R dv$, and the experimental value of Newton's constant at
ordinary scales, finds the coupling constant to be
\be \kappa_0(M_Z)=\sqrt{8\pi G}~,
\ee
thus,
\be
1/\kappa_0\sim 2.43\times
10^{18}\ {\rm GeV}~.
\ee
%                           √

Finally, the noncommutative geometry  approach to unification does
not  provide any  explanation of  the number  of  generations, neither
gives any constraints on the values of the Yukawa couplings.

\vskip.5truecm 

I believe there are two, distinct and highly nontrivial, avenues along
which the noncommutative geometry community has to make further
progress. Firstly, to include higher order corrections to the spectral
action and secondly, to find a noncommutative space, valid at
unification scale, whose limit is the {\sl almost} commutative
geometry ${\cal M}\times {\cal F}$ we have considered
insofar. Succeeding in these directions, may cure the small deviations
found between the predictions of the SM derived from the
noncommutative spectral geometry and the experimental values.

\section{Cosmological consequences of the NCG spectral action}

I will be using conventions in which the signature is $(-,+,+,+)$ and
the Ricci tensor is defined~\footnote{In General Relativity such
  choices are merely conventions, which are relatively unimportant; in
  the NCG approach we are following here the situation is however very
  different.}  as $R_{\mu\nu} = R^\rho\phantom{}_{\mu\nu\rho}$, with
$R_{\mu\nu\rho}\phantom{}^\sigma\omega_\sigma = \big[
  \bigtriangledown_\mu , \bigtriangledown_\nu \big] \omega_\rho$.  

The Lorentzian version of the gravitational part of the asymptotic
formula for the bosonic sector of the NCG spectral action, including
the coupling between the Higgs field and the Ricci curvature scalar,
reads~\cite{ccm}
\beq\label{eq:1.5} {\cal S}_{\rm grav}^{\rm L} = \int \left(
%\frac{1}{16\pi G} R 
\frac{1}{2\kappa_0^2} R + \alpha_0
C_{\mu\nu\rho\sigma}C^{\mu\nu\rho\sigma} + \tau_0 R^\star
R^\star\right.  \nonumber\\ -\left.  \xi_0 R|{\bf H}|^2 \right)
\sqrt{-g} \ d^4 x~. \eeq
The equations of motion arising from Eq.~(\ref{eq:1.5})
read~\cite{Nelson:2008uy}
\beq\label{eq:EoM2} R^{\mu\nu} - \frac{1}{2}g^{\mu\nu} R +
\frac{1}{\beta^2} \delta_{\rm cc}\left[
  2C^{\mu\lambda\nu\kappa}_{;\lambda ; \kappa} +
  C^{\mu\lambda\nu\kappa}R_{\lambda \kappa}\right]\nonumber\\ = 
%\ 8\pi G
\kappa_0^2 \delta_{\rm cc}T^{\mu\nu}_{\rm matter}~, \eeq
where $\beta^2$ and $\delta_{\rm cc}$ are defined as 
\be
\beta^2 \equiv -\frac{1}{4\kappa_0^2 \alpha_0}~,
\ee
and
\be
\delta_{\rm cc}\equiv[1-2\kappa_0^2\xi_0{\bf H}^2]^{-1}~,
\ee
respectively.

\subsection{Low energy regime}

Let me first concentrate on the low energy weak curvature regime,
where the nonminimal coupling term between the background geometry and
the Higgs field is small and can be safely neglected. This implies
that $\delta_{\rm cc}=1$.  For a
Friedmann-Lema\^{i}tre-Robertson-Walker (FLRW) space-time, the Weyl
tensor vanishes, hence the NCG corrections to the Einstein equation
vanish~\cite{Nelson:2008uy}.  This result has important
consequences. Since the NCG corrections vanish for FLRW cosmologies
and Schwarzschild solutions, it is difficult to place restrictions on
$\beta^2$ (or equivalently on $\alpha_0$, or on $f_0$), via cosmology
or solar-system tests.  Note that the best constraint on, different
{\it ad hoc}, curvature squared terms is obtained from measurements
of the orbital precession of Mercury, imposing the rather weak lower
bound~\cite{Stelle} $\beta_{R^2} \geq 3.2\times 10^{-9} {\rm m}^{-1}$.  Since
this constraint was found for terms of different form (but of the same
order) to the Weyl term appearing in the NCG spectral action, it does
not necessarily hold here.

We have constrained $\beta$ within the NCG spectral action context in
Ref.~\cite{wjm2}.  This constraint is very important, since by
imposing a lower limit to $\beta$, we actually set an upper limit to
the moment $f_0$ of the cut-off function used to define the spectral
action. Since $f_0$ can be used to specify the initial conditions on
the gauge couplings (see Eq.~(\ref{relation_g-f})), a constraint on
$\beta$ corresponds to a restriction on the particle physics at
unification. I will  briefly summarize how this constraint has been
obtained~\cite{wjm1,wjm2}.

Consider linear perturbations around a Minkowski background metric in
the synchronous gauge. The perturbed metric reads
\be g_{\mu\nu} = {\rm diag} \left( \{a(t)\}^2 \left[
  -1,\left(\delta_{ij} + h_{ij}\left(x\right)\right) \right]\right)~,
\ee
where $a(t)$ is the cosmological scale factor. Since we only consider
a flat background, $a(t)=1$ and $\dot a\equiv da/dt=0$. Note that the
remaining gauge freedom can be completely fixed by setting ${\bf
  \nabla}_i h^{ij}=0$.

The linearized equations of motion derived from the NCG spectral
action for such perturbations read~\cite{wjm1}
\be\label{eq:1} \left( \Box - \beta^2 \right) \Box h^{\mu\nu} =
\beta^2 \frac{16\pi G}{c^4} T^{\mu\nu}_{\rm matter}~, \ee 
where $T^{\mu\nu}_{\rm matter}$ is taken to lowest order in
$h^{\mu\nu}$. This implies that it is independent of $h^{\mu\nu}$
and satisfies the conservation equations
\be
\frac{\partial}{\partial x^\mu} T^\mu_{\ \nu}=0~.
\ee
Note that $\beta$ plays the r\^ole of a mass and hence has to be real
and positive, thus $\alpha_0 <0$.  For $\alpha_0>0$, as it has been
explicitly shown in Ref.~\cite{wjm1}, the gravitational waves evolve
according to a Klein-Gordon like equation with a tachyonic mass, and
hence the background, which has been considered to be a Minkowski
space, is unstable.  One hence has to restrict to $\alpha_0<0$ for
Minkowski space to be a (stable) vacuum of the theory.

Let us study the energy lost to gravitational radiation by orbiting
binaries. In the far field limit, $|{\bf r}| \approx |{\bf r} - {\bf
  r}'|$ (where ${\bf r}$ and ${\bf r}'$ stand for the locations of the
observer and emitter, respectively), the spatial components
of the general first order solution for a perturbation against a
Minkowski background read~\cite{wjm1}
\beq\label{eq:4} h^{ik}\left( {\bf r},t\right) \approx \frac{2G
  \beta}{3c^4} \int_{-\infty}^{t-\frac{1}{c}|{\bf r}|} \frac{d
  t'}{\sqrt{c^2\left( t-t'\right)^2 - |{\bf r}|^2} }\nonumber\\ \times
          {\cal J}_1 \left( \beta\sqrt{c^2\left( t-t'\right)^2 - |{\bf
              r}|^2}\right) \ddot{D}^{ik}\left(t'\right)~, \eeq
with ${\cal J}_1$ is a Bessel function of the first kind, in terms of
the quadrupole moment,
\be
D^{ik}\left(t\right) \equiv \frac{3}{c^2}\int  
x^i x^k T^{00}({\bf r},t) \ d{\bf r}~.
\ee
As one can easily check from Eq.~(\ref{eq:EoM2}), the theory reduces
to that of General Relativity in the $\beta\rightarrow \infty$ limit
and one can reproduce the familiar result for a massless graviton. For
finite $\beta$ however, gravitational radiation contains both massive
and massless modes, both of which are sourced from the quadrupole
moment of the system.

Consider a binary pair of masses $m_1, m_2$ in a circular (for
simplicity) orbit in the $(xy)$-plane. The rate of energy loss
from such a system, in the far field limit, is
\be\label{eq:energy} -\frac{{\rm d} {\cal E}}{{\rm d}t} \approx
\frac{c^2}{20G} |{\bf r}|^2 \dot{h}_{ij} \dot{h}^{ij}~,  \ee
with~\cite{wjm1}
\beq
&&\dot{h}^{ij}\dot{h}_{ij}= \frac{128\mu^2|\rho|^4 \omega^6 G^2
  \beta^2}{c^8}\nonumber\\
&&~~~~~~~~ \times \left[ f_{\rm c}^2\left(\beta|{\bf
    r}|,\frac{2\omega}{\beta c}\right) + f_{\rm s}^2\left(\beta|{\bf
    r}|,\frac{2\omega}{\beta c}\right)\right]~, \eeq
where we have defined the functions:
\beq\label{eq:f1}
 f_{\rm s}\left( x,z\right) &\equiv& \int_0^\infty
\frac{d s}{\sqrt{s^2 + x^2}} {\cal J}_1\left(s\right) \sin
\left(z\sqrt{ s^2 + x^2} \right)~,\nonumber\\
\label{eq:f2}
f_{\rm c}\left( x,z\right) &\equiv&
\int_0^\infty \frac{d s}{\sqrt{s^2 + x^2}} {\cal
  J}_1\left(s\right) \cos \left(z\sqrt{ s^2 + x^2} \right)~;\nonumber
\eeq
$\omega$ stands for the orbital frequency, which for the system under
consideration is a constant given by
\be \omega = |\rho|^{-3/2} \sqrt{ G\left( m_1 + m_2\right)}~,  \ee
with $|\rho|$ the magnitude of the separation vector between the two
bodies.

The integrals in Eq.~(\ref{eq:f1}), which can be easily evaluated for
$z<1$ and $z>1$, exhibit a strong resonance behavior at $z=1$, which
corresponds to the critical frequency~\cite{wjm1}
\be
\label{critical}
2\omega_{\rm c} =\beta c~,
\ee
around which strong deviations from the familiar results of General
Relativity are expected. This critical (maximum) frequency comes from
the natural length scale (given by $\beta^{-1}$) in the NCG theory, at
which noncommutative geometry effects become dominant.

For $z<1$ and $z>1$, the functions in Eq.~(\ref{eq:f1}) can be
evaluated numerically and fitted to an explicit functional
form~\cite{wjm1}. It can be then easily checked that for $\omega<
\omega_{\rm c}$, the $\beta \rightarrow \infty$ limit reproduces the
General Relativity result, as it should. Since this is not the case
for the $\omega>\omega_{\rm c}$ case, we consider the critical
frequency as the maximum one. Any deviation from the standard result
is suppressed by the distance to the source, at least for orbital
frequencies small compared to $\beta c$.

For the physically interesting case of $\omega<\omega_{\rm c}$, even
though the amplitude of the deviation from the standard result is
small, due to the $1/|{\bf r}|$ suppression, one should notice the
existence of a critical frequency $\omega_{\rm c}$ and the oscillatory
nature of the rate of flux of gravitational radiation with changing
distances and changing frequencies.

The form of the gravitational radiation from binary systems can be now
used to constrain $\beta$. Since we have considered only circular
binary orbits, we only need to know the orbital frequency and the
distance to the binary system.  Several binary pulsars, for which the
rate of change of the orbital frequency has been well characterized,
and the predictions of General Relativity agree with the data to high
accuracy have been used in Ref.~\cite{wjm2}.  The parameter $\beta$ has
been then restricted by requiring that the magnitude of deviations
from General Relativity be less than this uncertainty.  Using these
data and requiring that $\omega < \omega_{\rm c}$, we
found~\cite{wjm2}
\be 
\label{constr-beta}
\beta > 7.55\times 10^{-13}~{\rm m}^{-1}~.
\ee
Even though this observational constraint may seem weak, it is
comparable to (but larger than) existing constraints on similar, {\sl
  ad hoc}, additions to General Relativity, and one expects that it
will rapidly be improved as more binary pulsars are discovered and the
observations of existing systems improve.  

Due to the large distances to these binary systems, this constraint,
Eq.~(\ref{constr-beta}), is almost exactly due to $\beta > 2\omega
/c$. Thus, the strongest constraint comes from systems with high
orbital frequencies.  Future observations of rapidly orbiting
binaries, relatively close to the Earth, could thus improve this
constraint by many orders of magnitude.  Nevertheless, it is indeed
remarkable that the value of the Weyl squared coupling in the bosonic
action could be constrained~\cite{wjm2} via astrophysical data.

\vskip1.truecm

Let me go back to the background equations.  In order for the
corrections to Einstein's equations to be apparent at leading order,
i.e. at the level of the background, one needs to consider
anisotropic models. As an example, calculate the modified Friedmann
equation for the Bianchi type-V model, for which the space-time
metric in Cartesian coordinates reads
\be g_{\mu\nu} = {\rm diag} \left[ -1,\{a_1(t)\}^2e^{-2nz} ,
  \{a_2(t)\}^2e^{-2nz}, \{a_3(t)\}^2 \right]~, \ee
where $a(t)$, $b(t)$ and $c(t)$ are, in general, arbitrary functions
and $n$ is an integer. 

Writing down the modified Friedmann equation, we have
found~\cite{Nelson:2008uy} that the correction terms come in two
types. The first one contains terms which are fourth order in time
derivatives. Hence for the slowly varying functions, usually used in
cosmology, they can be taken to be small corrections. The second one
occurs at the same order as the standard Einstein-Hilbert
terms. However, it is proportional to $n^2$ and hence vanishes for
homogeneous versions of Bianchi type-V. Thus, although anisotropic
cosmologies do contain corrections due to the additional NCG terms in
the action, they are typically of higher order.  Inhomogeneous models
do contain correction terms that appear on the same footing as the
original (commutative) terms. In conclusion, the corrections to
Einstein's equations are present only in inhomogeneous and anisotropic
space-times.

\subsection{High energy regime}

At energies approaching the Higgs scale, the nonminimal coupling of
the Higgs field to the curvature can no longer be neglected, leading
to corrections even for background cosmologies. To understand the
effects of these corrections let us neglect the conformal term in
Eq.~(\ref{eq:EoM2}), i.e. set $\beta=0$. The equations of motion then
become~\cite{Nelson:2008uy}
\be R^{\mu\nu} - \frac{1}{2}g^{\mu\nu}R =
\kappa_0^2\left[\frac{1}{1-\kappa_0^2 |{\bf H}|^2/6}\right] T^{\mu\nu}_{\rm
  matter}~. \ee 
Hence, the $|{\bf H}|$ leads to an effective gravitational constant.

Alternatively, consider the effect of this term on the equations of
motion for the Higgs field in a constant gravitational field.  For
constant curvature, the self interaction of the Higgs field is
increased, as one can easily see from~\cite{Nelson:2008uy},
\be -\mu_0 |{\bf H}|^2 \rightarrow -\left( \mu_0 + \frac{R}{12}\right)
|{\bf H}|^2~.  \ee

\vskip1.truecm 

The nonminimal coupling between the Higgs field and the Ricci
curvature may turn out to be particularly useful in early universe
cosmology~\cite{Nelson:2009wr,mmm}.  Such a coupling has been
introduced {\sl ad hoc} in the literature, in an attempt to drive
inflation through the Higgs field.  However, the coupling constant
between the scalar field and the background geometry is not a free
parameter which could be tuned to achieve a successful inflationary
scenario, it should be instead dictated by the underlying theory.

Let me write again the Gravity-Higgs sector of the asymptotic
expansion of the spectral action, in Lorentzian
signature~\footnote{Since we are dealing with a FLRW metric, the Weyl
  tensor vanishes; the nondynamical term is also neglected.},
\be
\begin{split}
S^{\rm
  L}_{\rm GH}=\int\Big[\frac{1-2\kappa_0^2\xi_0
    H^2}{2\kappa_0^2}R 
\\
-\frac{1}{2}(\nabla  H)^2- V(H)\Big] \sqrt{-g}\  d^4x~,
\end{split}
\ee
where 
\be\label{higgs-pot}
V(H)=\lambda_0H^4-\mu_0^2H^2~,
\ee
with $\mu_0$ and $\lambda_0$ subject to radiative corrections as
functions of energy.  For large enough values of the Higgs field, the
renormalized value of these parameters must be calculated, while the
running of the top Yukawa coupling and the gauge couplings must be
evolved simultaneously.

At high energies the mass term is sub-dominant, and can be neglected
(only the first term in Eq.~(\ref{higgs-pot}) survives).  For each
value of the top quark mass, there is a value of the Higgs mass where
the effective potential is on the verge of developing a metastable
minimum at large values of the Higgs field and the Higgs potential is
locally flattened~\cite{mmm}.  Note that since the region where the
potential becomes flat is narrow, slow-roll must be very slow, in
order to provide a sufficiently long period of quasi-exponential
expansion, necessary to solve the shortcomings of the standard Hot Big
Bang cosmological model. Besides the slow-roll parameters, denoted by
$\epsilon$ and $\eta$, which may be slow enough to get sufficient
e-folds, the amplitude of density perturbations $\Delta_\mathcal{R}^2$
in the Cosmic Microwave Background must be in agreement with the
measured one. Inflation predicts that at horizon crossing (denoted by
stars), the amplitude of density perturbations is related to the
inflaton potential through
%q
\be \left(\frac{V_*}{\epsilon_*}\right)^{\frac14}
=2\sqrt{3\pi}\ m_\text{Pl}\ \Delta_\mathcal{R}^\frac12~,
\ee
where $\epsilon_*\leq1$.  Its value, as measured by
WMAP7~\cite{Larson:2010gs}, requires
\be \left(\frac{V_*}{\epsilon_*}\right)^{\frac14}
=(2.75\pm0.30)\times 10^{-2}\ m_\text{Pl}\,~,\label{eq:cobe} \ee
where $m_\text{Pl}$ stands for the Planck mass.  

In Ref.~\cite{mmm} we have calculated the renormalization of the Higgs
self-coupling up to two-loops and then constructed an effective
potential which fits the renormalization group improved potential
around the flat region.  We have found~\cite{mmm} that around the
plateau (the minimum of the potential), there is a very good analytic
fit to the Higgs potential, which takes the form
\beq
V^\text{eff}&=&\lambda_0^\text{eff}(H)H^4\nonumber\\
&=&[a\ln^2(b\kappa H)+c] H^4~,
\eeq
where the parameters $a, b$ are found to relate to the low energy
values of top quark mass $m_{\rm t}$ as~\cite{mmm}
\begin{align}
a(m_\text{t})&=4.04704\times10^{-3}-4.41909\times10^{-5}
\left(\frac{m_\text{t}}{\text{GeV}}\right)\nonumber\\ &\quad
+1.24732\times10^{-7}\left(\frac{m_\text{t}}{\text{GeV}}\right)^2~,
\nonumber\\ 
b(m_\text{t})&=\exp{\left[-0.979261
\left(\frac{m_\text{t}}{\text{GeV}}-172.051\right)\right]}~.
\end{align}
The third parameter, $c=c(m_\text{t},m_\phi)$, encodes the appearance
of an extremum and depends on the values for top quark mass and Higgs
mass.  An extremum occurs if and only if $c/a\leq 1/16$, the
saturation of the bound corresponding to a perfectly flat region.  
It is convenient to write $c=[(1+\delta)/16]a$, where $\delta=0$
saturates the bound below which a local minimum is formed.  

Note that the above strictly holds for the case of minimal coupling,
whereas in NCG we have a small nonminimal coupling, $\xi_0=1/12$.  The
corrections due to conformal coupling to the potential imply that
flatness does not occur at $\delta=0$ anymore but for fixed values of
$\delta$ depending on the value of the top quark mass. More precisely,
for inflation to occur via this mechanism, the top quark mass fixes
the Higgs mass extremely accurately.  Scanning through parameter space
it emerges that sufficient $e$-folds are indeed generated provided a
suitably tuned relationship between the top quark mass and the Higgs
mass holds~\cite{mmm}.

The exhaustive study of Ref.~\cite{mmm} has shown that while the Higgs
potential can lead to the slow-roll conditions being satisfied once
the running of the self-coupling at two-loops is included, the
constraints imposed from the CMB data make the predictions of such a
scenario incompatible with the measured value of the top quark mass.

Finally, running of the gravitational constant and corrections by
considering the more appropriate de\,Sitter, instead of a Minkowski,
background do not improve substantially the realization of a
successful inflationary era~\cite{mmm}.

\vskip 1.truecm

The NCG Spectral Action provides in addition to the Higgs field,
another (massless) scalar field~\cite{ali-sigma}, denoted by $\sigma$,
which is unlike all other fields in the theory, such as the Higgs
field and gauge fields. Note that $\sigma$ does not exhibit a coupling
to the matter sector.

Including this field, the cosmologically relevant terms in the Wick
rotated action read~\cite{ali-sigma}
\be
\begin{split}
S=\int\left[\frac{1}{2\kappa^2}
R - \xi_ H R H^2 - \xi_\sigma R \sigma^2 \right.\\ 
\left.-\frac{1}{2}(\nabla H)^2  -\frac{1}{2}(\nabla \sigma)^2 
- V(H,\sigma)\right]\ \sqrt{-g}\ d^4x~,
\end{split}
\ee
where
\be
V(H,\sigma)=\lambda_H H^4-\mu_H^2H^2+\lambda_\sigma
\sigma^4+\lambda_{ H\sigma}|H|^2\sigma^2~.
\ee
The constants are related to the underlying parameters as
follows:
\begin{align}
\xi_H &=\frac{1}{12}~~~~~~~~~~~~~~,  &\xi_\sigma &=\frac{1}{12}\\
\lambda_H &=\frac{\pi^2\mathfrak{b}}{2f_0\mathfrak{a}^2}~~~~~~~~~~, 
&\lambda_\sigma &= \frac{\pi^2\mathfrak{d}}{f_0\mathfrak{c}^2}\\
\mu_H &=2\Lambda^2\frac{f_2}{f_0}~~~~~~~~~~,
&\lambda_{H\sigma}&=\frac{2\pi^2\mathfrak{e}}{a\mathfrak{c}f_0}~.
\end{align}
Unfortunately, neither the $\sigma$ field can lead to a {\sl
  successful} slow-roll inflationary era, if the coupling values are
conformal~\cite{mmm}.  

\section{Conclusions}

Noncommutative spectral geometry is a beautiful mathematical
construction which offers an elegant and purely geometric
interpretation of the Standard Model of electroweak and strong
interactions. 

According to this proposal, the geometry near the Planck energy scale is the
tensor product of an internal discrete geometry for the SM and a
continuous geometry for space-time.  The unification is based on the
symplectic unitary group in Hilbert space and on the spectral
action. 

The NCG approach yields all the detailed structure of the SM with a
very little input, while it provides predictions at unification scale.
In particular, besides the familiar predictions for the gauge
couplings in agreement with GUTs theories, the model predicts the
Higgs scattering parameter and the sum of the squares of Yukawa
couplings.

Since the model lives by construction at unification scale, it
provides an excellent framework to address early universe cosmological
questions, while astrophysical issues are more difficult to be dealt.

In this contribution, I have first summarized the main mathematical
elements of noncommutative spectral geometry and then discussed
briefly some of its phenomenological and cosmological consequences.


\begin{thebibliography}{10}
%
\bibitem{ncg-book1} A.\ Connes, {\sl Noncommutative Geometry},
  Academic Press, New York (1994).

\bibitem{ncg-book2}  A.\ Connes and M.~Marcolli, {\sl Noncommutative Geometry,
 Quantum Fields and Motives}, Hindustan Book Agency, India (2008).

\bibitem{ccm} A.~H.~Chamseddine, A.~Connes and M.~Marcolli,
%``Gravity and the standard model with neutrino mixing,''
  Adv.\ Theor.\ Math.\ Phys.\  {\bf 11}, 991 (2007)
  [arXiv:hep-th/0610241].

\bibitem{ac2006}
A.\ Connes, JHEP {\bf 0611},081 (2006) [arXiv:hep-th/0608226].

\bibitem{fdp} J.\ Barrett, J.\ Math.\ Phys.\ {\bf 48}, 012303 (2007).

\bibitem{cc0706}
 A.~H.~Chamseddine, A.~Connes,
%WHY THE STANDARD MODEL
J.\ Geom.\ Phys.\ {\bf 58}, 38 (2008)
[arXiv:0706.3688 [hep-th]].

\bibitem{Chamseddine:2007ia}
A.~H.~Chamseddine and A.~Connes,
%``Conceptual Explanation for the Algebra in the Noncommutative Approach to
%the Standard Model,'' 
Phys.\ Rev.\ Lett.\ {\bf 99}, 191601 (2007)
  [arXiv:0706.3690 [hep-th]].

\bibitem{cc2005-dilaton}
A.~H.~Chamseddine and A.~Connes,
%Scale Invariance in the Spectral Action
J.\ of Math.\ Phys.\ {\bf 47}, 063504 (2006)͒
 [arXiv:hep-th/0512169].

\bibitem{sdw-coeff}
 A.~H.~Chamseddine and A.~Connes, J.~ Math.~Phys. {\bf 47}, 063504 (2006).

\bibitem{ac1996} A.~H.~Chamseddine and A.~Connes,
Phys.\ Rev.\ Lett.\ {\bf 77}, 4868 (1996).

\bibitem{ac1997} A.~H.~Chamseddine and A.~Connes,
  Comm.\ Math.\ Phys.\ {\bf 186}, 731 (1977).

\bibitem{nonpert}
 A.~H.~Chamseddine and A.~Connes, Comm.\ Math.\ Phys.\ {\bf 293}, 867 (2010)
arXiv:0812.0165 [hep-th].

%\cite{Nelson:2008uy}
\bibitem{Nelson:2008uy}
  W.~Nelson and M.~Sakellariadou,
  %``Cosmology and the Noncommutative approach to the Standard Model,''
  Phys.\ Rev.\ D {\bf 81}, 085038 (2010)
  [arXiv:0812.1657 [hep-th]].

%\cite{Nelson:2009wr}
\bibitem{Nelson:2009wr}
  W.~Nelson and M.~Sakellariadou,
  %``Natural inflation mechanism in asymptotic noncommutative geometry,''
  Phys.\ Lett.\  B {\bf 680}, 263 (2009)
  [arXiv:0903.1520 [hep-th]].
  %%CITATION = PHLTA,B680,263;%%

%\cite{Marcolli:2009in}
\bibitem{Marcolli:2009in}
  M.~Marcolli and E.~Pierpaoli,
  %``Early Universe models from Noncommutative Geometry,''
  arXiv:0908.3683 [hep-th].
  %%CITATION = ARXIV:0908.3683;%%

%\cite{mmm}
\bibitem{mmm}
M.~Buck, M.~Fairbairn and M.~Sakellariadou,
%``Inflation within models with Conformal Coupling between the
%  Scalar field and the Ricci Curvature: An application to the
%  Noncommutative Spectral Action,'' 
Phys.\ Rev.\ D {\bf 82}, 043509 (2010)
[arXiv:1005.1188 [hep-th]].

\bibitem{wjm1}
W.~Nelson, J.~Ochoa and M.~Sakellariadou,
[arXiv:1005.4276 [hep-th]].

\bibitem{wjm2}
W.~Nelson, J.~Ochoa and M.~Sakellariadou,
[arXiv:1005.4279 [hep-th]].

\bibitem{Buchbinder:1989bt}
I.~L. Buchbinder, S.~D. Odintsov and I.~M. Lichtzier,
Class.\ Quant.\ Grav.\ {\bf 6}, 605 (1989).

\bibitem{Yoon:1996yr}
Youngsoo Yoon and Yongsung Yoon,
Int.\ J.\ Mod.\ Phys.\ A {\bf 12}, 2903 (1997.

\bibitem{mpt}
M.~Marcolli E.~Pierpaoli and K.\ Teh,
[arXiv:1005.2256 [hep-th]].

%\cite{Stelle}
\bibitem{Stelle}
 K.~S.~Stelle,
 %``Classical Gravity with Higher Derivatives,''
 Gen. Rel. Grav. {\bf 9} 353 (1978).

\bibitem{Larson:2010gs}
D.~Larson et~al.,
[arXiv:1001.4758 [astro-ph]].

\bibitem{ali-sigma}
A.~H.~Chamseddine, 
[arXiv:0901.0577 [hep-th]].

\end{thebibliography}
\end{document}